\documentstyle[preprint,aps]{revtex}

\newcommand{\krc}{k r_c}
\newcommand{\krcp}{k r_c\pi}

\begin{document}

\setlength\baselineskip{20pt}

\preprint{\tighten\vbox{\hbox{CALT-68-2232}\hbox{hep-ph/9907447}}}

\title{Modulus Stabilization with Bulk Fields}

\author{Walter D. Goldberger\footnote{walter@theory.caltech.edu} and Mark B. Wise\footnote{wise@theory.caltech.edu}}
\address{\tighten California Institute of Technology, Pasadena, CA 91125}

\maketitle

{\tighten
\begin{abstract}
We propose a mechanism for stabilizing the size of the extra dimension in the Randall-Sundrum scenario.  The potential for the modulus field that sets the size of the fifth dimension is generated by a bulk scalar with quartic interactions localized on the two 3-branes.  The minimum of this potential yields a compactification scale that solves the hierarchy problem without fine tuning of parameters.
\end{abstract}}
\vspace{0.7in}
\narrowtext

\newpage

The Standard Model for strong, weak, and electromagnetic interactions based on the gauge group $SU(3)\times SU(2)\times U(1)$ has been extremely succesful in accounting for experimental observations.  However, it has several unattractive features that suggest new physics beyond that incorporated in this model.  One of these is the gauge hierarchy problem, which refers to the vast disparity between the weak scale and the Planck scale.  In the context of the minimal Standard Model, this hierarchy of scales is unnatural since it requires a fine tuning order by order in perturbation theory.  A number of extensions have been proposed to solve the hierarchy problem, notably technicolor\cite{tech} (or dynamical symmetry breaking) and low energy supersymmetry\cite{SUSY}.  

Recently, it has been suggested that large compactified extra dimensions may provide an alternative solution to the hierarchy problem\cite{xdim}.  In these models, the observed Planck mass $M_{Pl}$ is related to $M,$ the fundamental mass scale of the theory, by $M_{Pl}^2 = M^{n+2} V_n,$ where $V_n$ is the volume of the additional compactified dimensions.  If $V_n$ is large enough, $M$ can be of the order of the weak scale.  Unfortunately, unless there are several large extra dimensions, a new hierarchy is introduced between the compactification scale, $\mu_c=V_n^{-1/n},$ and $M.$

Randall and Sundrum~\cite{RS1} have proposed a higher dimensional scenario to solve the hierachy problem that does not require large extra dimensions.  This model consists of a spacetime with a single $S^1/Z_2$ orbifold extra dimension.  Three-branes with opposite tensions reside at the orbifold fixed points and together with a finely tuned cosmological constant serve as sources for five-dimensional gravity.  The resulting spacetime metric contains a redshift factor which depends exponentially on the radius $r_c$ of the compactified dimension:
\begin{equation}
\label{eq:metric}
ds^2 = e^{-2 \krc |\phi|}\eta_{\mu\nu} dx^\mu dx^\nu - r_c^2 d\phi^2,
\end{equation}
where $k$ is a parameter which is assumed to be of order $M$, $x^\mu$ are Lorentz coordinates on the four-dimensional surfaces of constant $\phi$, and $-\pi\leq \phi\leq\pi$ with $(x,\phi)$ and $(x,-\phi)$ identified.  The two 3-branes are located at $\phi=0$ and $\phi=\pi.$   A similar scenario to the one described in ref.~\cite{RS1} is that of Horava and Witten~\cite{HW}, which arises within the context of $M$-theory.  Supergravity solutions similar to Eq.~(\ref{eq:metric}) are presented in ref.~\cite{supergrav}. In ref.~\cite{verlinde}, it is shown how this model may be obtained from string theory compactifications.  

The non-factorizable geometry of Eq.~(\ref{eq:metric}) has several important consequences.  For instance, the four-dimensional Planck mass is given in terms of the fundamental scale $M$ by
\begin{equation}
M_{Pl}^2=\frac{M^3}{k}[1-e^{-2kr_c\pi}],
\end{equation}
so that, even for large $\krc,$ $M_{Pl}$ is of order $M.$  Because of the exponential factor in the spacetime metric, a field confined to the 3-brane at $\phi=\pi$ with mass parameter $m_0$ will have physical mass $m_0 e^{-\krcp}$ and for $kr_c$ around 12, the weak scale is dynamically generated from a fundamental scale $M$ which is on the order of the Planck mass.  Furthermore, Kaluza-Klein gravitational modes have TeV scale mass splittings and couplings\cite{RS2}.  Similarly, a bulk field with mass on the order of $M$ has low-lying Kaluza-Klein excitations that reside primarily near $\phi=\pi$ and hence, from a four-dimensional perspective, have masses on the order of the weak scale~\cite{us}.

In the scenario presented in ref.~\cite{RS1}, $r_c$ is associated with the vacuum expectation value of a massless four-dimensional scalar field.  This modulus field has zero potential and consequently $r_c$ is not determined by the dynamics of the model.  For this scenario to be relevant, it is necessary to find a mechanism for generating a potential to stabilize the value of $r_c.$  Here we show that such a potential can arise classically from the presence of a bulk scalar with interaction terms that are localized to the two 3-branes\footnote{Other proposals for stabilizing the $r_c$ modulus can be found in ref.~\cite{steinhardt}.}.  The minimum of this potential can be arranged to yield a value of $\krc\sim 10$ without fine tuning of parameters.   

Imagine adding to the model a scalar field $\Phi$ with the following bulk action
\begin{equation}
S_b={1\over 2}\int d^4 x\int_{-\pi}^\pi d\phi \sqrt{G} \left(G^{AB}\partial_A \Phi \partial_B \Phi - m^2 \Phi^2\right),
\end{equation}
where $G_{AB}$ with $A,B=\mu,\phi$ is given by Eq.~(\ref{eq:metric}).  We also include interaction terms on the hidden and visible branes (at $\phi=0$ and $\phi=\pi$ respectively) given by
\begin{equation}
S_h = -\int d^4 x \sqrt{-g_h}\lambda_h \left(\Phi^2 - v_h^2\right)^2,
\end{equation}
and
\begin{equation}
S_v = -\int d^4 x \sqrt{-g_v}\lambda_v \left(\Phi^2 - v_v^2\right)^2,
\end{equation}
where $g_h$ and $g_v$ are the determinants of the induced metric on the hidden and visible branes respectively.  Note that $\Phi$ and $v_{v,h}$ have mass dimension $3/2$, while $\lambda_{v,h}$ have mass dimension $-2.$ Kinetic terms for the scalar field can be added to the brane actions without changing our results.  The terms on the branes cause $\Phi$ to develop a $\phi$-dependent vacuum expectation value $\Phi(\phi)$ which is determined classically by solving the differential equation
\begin{eqnarray}
\label{eq:eom}
0 &=& -{1\over r_c^2}\partial_\phi\left(e^{-4\sigma}\partial_\phi\Phi\right)+m^2 e^{-4\sigma}\Phi + 4e^{-4\sigma}\lambda_v\Phi \left(\Phi^2 - v_v^2\right)\frac{\delta(\phi-\pi)}{r_c}\nonumber \\
& & \mbox{} + 4e^{-4\sigma}\lambda_h\Phi \left(\Phi^2 - v_h^2\right)\frac{\delta(\phi)}{r_c},
\end{eqnarray}
where $\sigma(\phi)=\krc |\phi|.$  Away from the boundaries at $\phi=0,\pi$, this equation has the general solution
\begin{equation}
\label{eq:soln}
\Phi(\phi) = e^{2\sigma}[A e^{\nu\sigma}+B e^{-\nu\sigma}],
\end{equation}
with $\nu=\sqrt{4+m^2/k^2}$.  Putting this solution back into the scalar field action and integrating over $\phi$ yields an effective four-dimensional potential for $r_c$ which has the form
\begin{eqnarray}
\label{eq:pot}
V_\Phi(r_c)&=&k(\nu+2)A^2 (e^{2\nu k r_c\pi}-1)+k(\nu-2)B^2(1-e^{-2\nu\krcp})\nonumber\\
& & \hbox{} + \lambda_v e^{-4\krcp}\left(\Phi(\pi)^2-v_v^2\right)^2+\lambda_h \left(\Phi(0)^2-v_h^2\right)^2.
\end{eqnarray}
The unknown coefficients $A$ and $B$ are determined by imposing appropriate boundary conditions on the 3-branes.  We obtain these boundary conditions by inserting Eq.~(\ref{eq:soln}) into the equations of motion and matching the delta functions:
\begin{eqnarray}
\label{eq:bc1} k\left[(2+\nu) A + (2-\nu)B\right] -2\lambda_h\Phi(0)\left[\Phi(0)^2 - v_h^2\right]=0,\\
\label{eq:bc2} k e^{2\krcp}\left[(2+\nu) e^{\nu\krcp} A + (2-\nu) e^{-\nu\krcp}B\right] +2\lambda_v\Phi(\pi)\left[\Phi(\pi)^2 - v_v^2\right]=0.
\end{eqnarray}

Rather than solve these equations in general, we consider the simplified case in which the parameters $\lambda_h$ and $\lambda_v$ are large.  It is evident from Eq.~(\ref{eq:pot}) that in this limit, it is energetically favorable\footnote{\tighten The configuration that has both VEVs of the same sign has lower energy than the one with alternating signs and therefore corresponds to the ground state.  Clearly, the overall sign is irrelevant.} to have $\Phi(0)=v_h$ and $\Phi(\pi)=v_v$.  Thus, from Eq.~(\ref{eq:soln}) we get for large $\krc$
\begin{eqnarray}
A &=& v_v e^{-(2+\nu)\krcp} - v_h e^{-2\nu\krcp},\\
B &=& v_h (1+e^{-2\nu\krcp}) - v_v e^{-(2+\nu)\krcp},
\end{eqnarray}
where subleading powers of $\exp (-\krcp)$ have been neglected.  Now suppose that $m/k\ll 1$ so that $\nu=2+\epsilon,$ with $\epsilon\simeq m^2/4k^2$ a small quantity.  In the large $\krc$ limit, the potential becomes
\begin{equation}
\label{eq:leading}
V_\Phi(r_c)= k\epsilon v_h^2 + 4ke^{-4\krcp}(v_v - v_h e^{-\epsilon\krcp})^2\left(1+\frac{\epsilon}{4}\right) - k\epsilon v_h e^{-(4+\epsilon)\krcp}(2 v_v - v_h e^{-\epsilon\krcp})
\end{equation}
where terms of order $\epsilon^2$ are neglected (but $\epsilon\krc$ is not treated as small).  Ignoring terms proportional to $\epsilon$, this potential has a minimum at 
\begin{equation}
\label{eq:min}
\krc = \left(\frac{4}{\pi}\right) \frac{k^2}{m^2} \ln\left[\frac{v_h}{v_v}\right].
\end{equation}
With $\ln (v_h/v_v)$ of order unity, we only need $m^2/k^2$ of order $1/10$ to get $\krc\sim 10.$  Clearly, no extreme fine tuning of parameters is required to get the right magnitude for $\krc.$  For instance, taking  $v_h/v_v=1.5$ and $m/k = 0.2$ yields $\krc\simeq 12.$

The stress tensor for the scalar field can be written as $T^{AB}_s=T^{AB}_{k}+T^{AB}_m,$ where for large $\krc$:
\begin{eqnarray}
T^{\phi\phi}_k &\simeq& -\frac{k^2}{2 r_c^2}\left[(4+\epsilon)(v_v-v_h e^{-\epsilon\krcp}) e^{-(4+\epsilon)(\krcp -\sigma)}-\epsilon v_h e^{-\epsilon\sigma}\right]^2,\\
T^{\mu\nu}_k   &\simeq& \frac{k^2}{2} e^{2\sigma}\eta^{\mu\nu} \left[(4+\epsilon)(v_v-v_h e^{-\epsilon\krcp}) e^{-(4+\epsilon)(\krcp -\sigma)}-\epsilon v_h e^{-\epsilon\sigma}\right]^2,
\end{eqnarray}
and 
\begin{eqnarray}
T^{\phi\phi}_m &\simeq& -\frac{2 k^2\epsilon}{r_c^2}\left[(v_v-v_h e^{-\epsilon\krcp}) e^{-(4+\epsilon)(\krcp-\sigma)} + v_h e^{-\epsilon\sigma}\right]^2,\\
T^{\mu\nu}_m   &\simeq& -2 k^2 e^{2\sigma}\eta^{\mu\nu} \epsilon \left[(v_v-v_h e^{-\epsilon\krcp}) e^{-(4+\epsilon)(\krcp-\sigma)} + v_h e^{-\epsilon\sigma}\right]^2.
\end{eqnarray}
As long as $v_h^2/M^3$ and $v_v^2/M^3$ are small, $T^{AB}_s$ can be neglected in comparison to the stress tensor induced by the bulk cosmological constant.  It is therefore safe to ignore the influence of the scalar field on the background geometry for the computation of $V(r_c)$.  A similar criterion ensures that the stress tensor induced by the bulk cosmological constant is dominant for $\krc\sim 1.$  

One might worry that the validity of Eq.~(\ref{eq:leading}) and Eq.~(\ref{eq:min}) requires unnaturally large values of $\lambda_h$ and $\lambda_v$.  We will check that this is not the case by computing the leading $1/\lambda$ correction to the potential.  To obtain this correction, we linearize Eq.~(\ref{eq:bc1}) and Eq.~(\ref{eq:bc2}) about the large $\lambda$ solution.  Neglecting terms of order $\epsilon$, the VEVs are shifted by 
\begin{eqnarray}
\delta\Phi(0) &=&\frac{k}{\lambda_h v_h^2} e^{-(4+\epsilon)\krcp}(v_v - v_h e^{-\epsilon\krcp}),\\
\delta\Phi(\pi) &=&-\frac{k}{\lambda_v v_v^2}(v_v - v_h e^{-\epsilon\krcp}),
\end{eqnarray}
and thus (neglecting subleading exponentials of $\krcp$)
\begin{eqnarray}
\delta A &=& -\frac{k}{\lambda_v v_v^2}e^{-(4+\epsilon)\krcp}(v_v - v_h e^{-\epsilon\krcp}), \\
\delta B &=&  e^{-(4+\epsilon)\krcp}(v_v - v_h e^{-\epsilon\krcp})\left[\frac{k}{\lambda_v v_v^2} + \frac{k}{\lambda_h v_h^2}\right].
\end{eqnarray}
Hence, the correction to the potential is 
\begin{equation}
\label{eq:corr}
\delta V_\Phi(r_c) = -\frac{4k^2}{\lambda_v v_v^2} e^{-4\krcp}(v_v - v_h e^{-\epsilon\krcp})^2.
\end{equation}
This has the same form as the leading $\epsilon\rightarrow 0$ behavior of Eq.~(\ref{eq:leading}) and therefore does not significantly affect the location of the minimum.

Note that the forms of the potentials in Eq.~(\ref{eq:leading}) and Eq.~(\ref{eq:corr}) are only valid for large $\krc$.  For small $\krc$, the potential becomes 
\begin{equation}
V_\Phi (r_c) = \frac{(v_v-v_h)^2}{\pi r_c}
\end{equation}
when terms of order $\epsilon$ and $1/\lambda$ are neglected.  The singularity as $r_c\rightarrow 0$ is removed by finite $\lambda$ corrections which become large for small $r_c$, and yield 
\begin{equation}
V_\Phi (0) = \frac{\lambda_h \lambda_v}{\lambda_h+ \lambda_v} \left(v_v^2 - v_h^2\right)^2.
\end{equation}  

In the scenario of Randall and Sundrum, the action is the sum of the five-dimensional Einstein-Hilbert action plus world-volume actions for the 3-branes:
\begin{equation}
\label{eq:RS}
S =\int d^4 x d\phi \sqrt{G}[-\Lambda + 2 M^3 R] - \int d^4 x \sqrt{-g_h} V_h - \int d^4 x \sqrt{-g_v} V_v \nonumber.
\end{equation}
For Eq.~(\ref{eq:metric}) to be a solution of the field equations that follow from Eq.~(\ref{eq:RS}), one must arrange $V_h=-V_v=24 M^3 k,$ where $\Lambda=-24 M^3 k^2$.  This amounts to having a vanishing four-dimensional cosmological constant plus an additional fine tuning which causes the $r_c$ potential to vanish.  However, imagine perturbing the 3-brane tensions by small amounts
\footnote{\tighten It has been noted that given the action in Eq.~(\ref{eq:RS}), changes in the relation between the brane tensions and the bulk cosmological constant result in bent brane solutions\cite{curved}.  It is possible that there are higher dimension induced curvature terms in the brane actions that make it energetically favorable for them to stay flat.  For $V_h=-V_v=24 M^3 k,$ Eq.~(\ref{eq:metric}) remains a solution to the field equations in the presence of such terms.}:
\begin{eqnarray}
V_h\rightarrow V_h+\delta V_h,\\
V_v\rightarrow V_v+\delta V_v.
\end{eqnarray}
As long as $|\delta V_h|$ and $|\delta V_v|$ are small compared to $-\Lambda/k,$ these shifts in the brane tensions induce the following potential for $r_c$
\begin{equation}
V_\Lambda (r_c) = \delta V_h + \delta V_v e^{-4\krcp}.
\end{equation}
For $\delta V_v$ small, the sum of potentials $V_\Phi(r_c) + V_\Lambda(r_c)$ has a minimum for large $\krc.$  The effective  four-dimensional cosmological constant can be tuned to zero by adjusting the value of $\delta V_h.$  For $\delta V_v < k\epsilon v_v^2$ the minimum is global, while for $\delta V_v > k\epsilon v_v^2$ the minimum is a false vacuum since $r_c\rightarrow \infty$ is a configuration of lower energy.

We have seen that a bulk scalar with a $\phi$-dependent VEV can generate a potential to stabilize $r_c$ without having to fine tune the parameters of the model (there is still one fine tuning associated with the four-dimensional cosmological constant, however).  This mechanism for stabilizing $r_c$ is a reasonably generic effect caused by the presence of a $\phi$-dependent vacuum bulk field configuration.  It may be worthwile to work out other features of the specific toy model presented here, such as the back reaction of the scalar field and shifts of the brane tensions on the spacetime geometry.  Also, the use of the large $\lambda$ limit was purely for convenience and the finite $\lambda$ case could be considered.  

With $r_c$ stabilized, the cosmology associated with this scenario should be standard for temperatures below the weak scale.  However, for temperatures above this scale, it will be different (see ref.~\cite{cosmo}) from the usual Friedmann cosmology.

The scenario presented in ref.~\cite{RS1} represents an attractive solution to the hierarchy puzzle.  However, it also has some features that are less appealing than the Standard Model with minimal particle content.  In this scenario, higher dimension operators are suppressed by the weak scale and unlike the Standard Model, where the suppression can be by powers of the GUT scale, there is no explanation for the smallness of neutrino masses and the long proton lifetime based simply on dimensional analysis.

We thank R. Sundrum for several useful conversations.  This work was supported in part by the Department of Energy under grant number DE-FG03-92-ER 40701.

\end{document}